\documentclass[12pt]{msml2020} 

\title[Quantum Ground States from Reinforcement Learning]{Quantum Ground States from Reinforcement Learning}
\usepackage{hyperref}
\usepackage{times, braket, cleveref, easy-todo}
\usepackage{graphicx}
\usepackage{tabulary}
\usepackage{comment}

\msmlauthor{\Name{Ariel Barr} \Email{arb225@cam.ac.uk}\\
 \Name{Willem Gispen} \Email{wg265@cam.ac.uk}\\
 \Name{Austen Lamacraft} \Email{al200@cam.ac.uk}\\
 \addr TCM Group, Cavendish Laboratory, University of Cambridge, J. J. Thomson Ave., Cambridge CB3 0HE, UK}

\DeclareMathOperator*{\E}{\mathbb{E}}
\newcommand*{\bv}{\vec{v}}
\newcommand*{\br}{\vec{r}}
\newcommand*{\bz}{\vec{z}}

\begin{document}

\maketitle

\begin{abstract}%
  Finding the ground state of a quantum mechanical system can be formulated as an optimal control problem. In this formulation, the drift of the optimally controlled process is chosen to match the distribution of paths in the Feynman--Kac (FK) representation of the solution of the imaginary time Schr\"odinger equation. This provides a variational principle that can be used for reinforcement learning of a neural representation of the drift. Our approach is a drop-in replacement for path integral Monte Carlo, learning an optimal importance sampler for the FK trajectories. We demonstrate the applicability of our approach to several problems of one-, two-, and many-particle physics.
\end{abstract}


\section{Introduction}\label{sec:intro}

Quantum mechanics takes place in infinite dimensional Hilbert space. Naturally, any numerical approach to solving the equations of quantum mechanics -- or any other physical system with a continuum description -- involves a finite truncation of this space. When we turn to \emph{many-body} quantum mechanics, the dimension of Hilbert space necessarily grows exponentially with the number of particles relative to the finite truncation used for a single particle. To be specific, if a single particle is described by a wavefunction $\psi(\vec{r})$ defined on a real-space grid of linear size $L$ (with $L^3$ points in three dimensions), the wavefunction of $N$ particles $\Psi(\vec{r}_1,\ldots,\vec{r}_N)$ is defined on a grid in $3N$ dimensions of $L^{3N}$ points.

Traditionally, this challenge has been dealt with by considering many body wavefunctions of restricted form. For example, the Hartree--Fock method employs factorized wavefunctions\footnote{We leave aside the issue of the statistics of indistinguishable particles for the moment.}
$$
\Psi(\vec{r}_1,\ldots,\vec{r}_N)=\psi_1(\vec{r}_1)\ldots \psi_N(\vec{r}_N).
$$
This reduces the memory cost to linear in the number of particles,\footnote{The computational complexity of the Hartree--Fock method scales as the cube of the number of basis functions used to represent the $\psi_i(\vec{r})$.} but represents a drastic simplification that performs especially poorly when the interaction between particles is strong. Over the years many \emph{post--Hartree--Fock} hand-crafted improvements of the many-body wavefunction have been introduced, including Jastrow factors, and the coupled cluster and configuration interactions methods (\cite{Foulkes:2001aa}).

The exponential growth in the complexity of many-body quantum mechanics closely parallels the curse of dimensionality encountered in computer vision and other traditional applications of machine learning. Therefore it is natural that neural methods that have recently proven successful in the latter domain be applied to quantum mechanical calculations.


\subsection{Deep Learning Approaches to Quantum Ground States} \label{sec:dlq}

Beginning with the work of \cite{Carleo:2017aa}, the past few years have seen numerous attempts to leverage the expressive power of deep networks to solve the quantum many-body problem.\footnote{There is earlier work on few-body quantum mechanics from before the deep learning era (\cite{Lagaris:1997aa}). A separate strand of work is concerned with efficient parameterization of interatomic potentials that have been calculated by other means, see e.g. (\cite{Behler:2016aa}).}

This initial work dealt with lattice models of spins that are a mainstay of quantum condensed matter physics, representing the wavefunction using a Restricted Boltzmann Machine. Since then neural representations of many body wavefunctions have multiplied to include lattice systems of fermions (\cite{Nomura:2017aa}) and bosons (\cite{Saito:2017aa}), particles in continuous space (\cite{Ruggeri:2018aa,Kessler:2019aa}), systems with symmetry (\cite{Choo:2018aa}), and the many-electron problem (\cite{Han:2019aa,Pfau:2019aa,Hermann:2019aa}).


All of the above-mentioned applications of neural networks to quantum mechanics have used the Schr\"odinger picture, in which a representation is sought of the wavefunction. There are several mathematically equivalent formulations of quantum mechanics that provide alternatives to the Schr\"odinger equation, and hence potential alternative routes for the application of machine learning. In this work we show how \emph{reinforcement learning} may be used to solve problems in quantum mechanics via the \emph{path integral} representation.

\subsection{Feynman--Kac Representation of the Ground State}

Feynman's path integral formulation of quantum mechanics starts from the Schr\"odinger equation (\cite{Feynman:1965aa})
\begin{equation}
  i\frac{\partial\psi(\vec{r},t)}{\partial t} = \left[H\psi\right](\vec{r},t),
\end{equation}
where for a single particle the Hamiltonian operator describing motion in a potential $V(\vec{r})$ has the form
\begin{equation}
  \left[H\phi\right](\vec{r}) = -\frac{1}{2}\nabla^2\phi(\vec{r}) + V(\vec{r})\phi(\vec{r}).
\end{equation}
The wavefunction at time $t_2>t_1$ can be expressed in terms of the wavefunction at time $t_1$ by
\begin{equation}
  \psi(\vec{r}_2,t_2) = \int d\vec{r}_1 \mathcal{K}(\vec{r}_2,t_2;\vec{r}_1,t_1)\psi(\vec{r}_1,t_1),
\end{equation}
where the \emph{propagator} $\mathcal{K}(\vec{r}_2,t_2;\vec{r}_1,t_1)$ is the Green's function of the Schr\"odinger equation
\begin{equation}
  \left(i\frac{\partial}{\partial t_2} - H_{\vec{r}_2}\right)\mathcal{K}(\vec{r}_2,t_2;\vec{r}_1,t_1) =i\delta(\vec{r}_1-\vec{r}_2)\delta(t_1-t_2).
\end{equation}
$\mathcal{K}(\vec{r}_2,t_2;\vec{r}_1,t_1)$ has the physical interpretation of the \emph{probability amplitude} to move from $\vec{r}_1$ at time $t_1$ to $\vec{r}_2$ at time $t_2$. The \emph{Born rule} then states that the square modulus of the probability amplitude gives the probability for this event.

The path integral is a representation of the propagator as a `sum over paths', written formally as
\begin{equation}\label{eq:real-time}
  \mathcal{K}(\vec{r}_2,t_2;\vec{r}_1,t_1) = \int_{\vec{r}(t_1)=\vec{r}_1 \atop \vec{r}(t_2)=\vec{r}_2} \mathcal{D}\vec{r}(t)\exp\left(i\int_{t_1}^{t_2} L(\vec{r},\dot{\vec{r}})dt\right)
\end{equation}
where $L(\vec{r},\vec{v}) = \frac{1}{2}\vec{v}^2 - V(\vec{r})$ is the classical Lagrangian function of the system, and the domain of the integral is all paths satisfying the stated endpoint conditions.

Despite being successfully wielded by physicists for decades Feynman's original idea has never found rigorous mathematical formulation due to the difficulties of defining a suitable measure on the path space. \cite{Kac:1949aa} discovered, however, that a fully rigorous path integral formula exists for the heat-type equations
\begin{equation}\label{eq:im-time}
  \frac{\partial\psi(\vec{r},t)}{\partial t} = -\left[H\psi\right](\vec{r},t),
\end{equation}
also known as the imaginary time Schr\"odinger equation. Moving to imaginary time makes the exponent in \eqref{eq:real-time} real. Kac observed that the part of the exponent arising from the kinetic energy could be interpreted as a measure on Brownian paths, leading to the \emph{Feynman--Kac (FK) formula} for the solution of \eqref{eq:im-time}
\begin{equation}\label{eq:FK}
  \psi(\vec{r}_2,t_2) =  \E_{\vec{r}(t_2)=\vec{r}_2}\left[\exp\left(-\int_{t_1}^{t_2}V(\vec{r}(t))dt\right)\psi(\vec{r}(t_1),t_1)\right],
\end{equation}
where the expectation is over Brownian paths finishing at $\vec{r}_2$ at time $t_2$. In this way quantum mechanics is brought into the realm of stochastic processes, albeit in `imaginary time'. Whilst this formulation is therefore not of direct utility in studying quantum dynamics, it provides a very useful tool for studying \emph{ground states}. This is because the propagator $K(\vec{r}_2,t_2;\vec{r}_1,t_1)$ for \eqref{eq:im-time} has a spectral representation in terms of the eigenfunctions $\varphi_n$ and eigenenergies $E_n$ of the time independent Schr\"odinger equation $H\varphi_n = E_n\varphi_n$ as
\begin{align}\label{eq:long-time}
  K(\vec{r}_2,t_2;\vec{r}_1,t_1) &= \sum_n \varphi_n(\vec{r}_2)\varphi^*_n(\vec{r}_1)e^{-E_n(t_2-t_1)}\\
  &\longrightarrow \varphi_0(\vec{r}_2)\varphi^*_0(\vec{r}_1)e^{-E_0(t_2-t_1)} \qquad \text{ as } t_2-t_1\to\infty.
\end{align}
Thus, as $t_2-t_1\to\infty$, only the ground state contributes.

The FK formula defines a new path measure $\mathbb{P}_\text{FK}$ that differs from the Brownian measure $\mathbb{P}_0$ by the Radon--Nikodym derivative
\begin{equation}\label{eq:RN}
  \frac{d\mathbb{P}_\text{FK}}{d\mathbb{P}_0} = \mathcal{N}\exp\left(-\int_{t_1}^{t_2}V(\vec{r}(t))dt\right)
\end{equation}
where $\mathcal{N}$ is a normalization factor. Intuitively, \eqref{eq:RN} describes paths that spend more time in the attractive regions of the potential ($V(\vec{r})<0$) and less time in the repulsive regions ($V(\vec{r})>0$). It is natural to conjecture that if $t_2=-t_1=T/2$ with $T\to\infty$, the distribution of $\vec{r}(0)$ under this measure  coincides with the ground state probability distribution  $|\varphi_0(\vec{r})|^2$ from the Born rule. To see that this is the case, consider a path that passes through $(\vec{r}_-,-T/2)$, $(\vec{r},0)$ and $(\vec{r}_+,T/2)$ for some arbitrary initial and final point $\vec{r}_\pm$. The overall propagator is then
\begin{equation}
  K(\vec{r}_+,T/2;\vec{r},0)K(\vec{r},0;\vec{r}_-,-T/2;)\sim  |\varphi_0(\vec{r})|^2\varphi_0(\vec{r}_+)\varphi^*_0(\vec{r}_-)e^{-E_0T}.
\end{equation}
Apart from a normalization factor that depends on $\vec{r}_\pm$ and $T$,\footnote{In particular, we see that $\mathcal{N}\sim e^{E_0 T}$ in the long time limit.} this is just the expected ground state distribution $|\varphi_0(\vec{r})|^2$. Thus, the ability to sample from the FK measure for long trajectories would also allow us to sample from the ground state distribution.

\subsection{Quantum Mechanics and Optimal Control}

Long before Feynman, the path measure was studied for finite $T$ and the case of a free particle ($V(\vec{r})=0$) by \cite{Schrodinger:1931aa,Schrodinger:1932aa} in an early exploration of the connection between his equation and the heat equation. Schr\"odinger sought the path measure that interpolates between given marginal distributions at the initial and final times. The generalizations of this question are now known as the \emph{Schr\"odinger problem}: see \cite{Leonard:2014aa} for an insightful recent review with a historical survey. One of the major conclusions of these works is that the path measure is \emph{Markov} (\cite{Jamison:1974aa}) and that paths $\vec{r}(t)$ satisfy a stochastic differential equation (SDE)
\begin{equation}\label{eq:sde}
  d\vec{r}_t = d\vec{B}_t + \vec{v}(\vec{r}_t,t)dt,
\end{equation}
where $\vec{B}_t\in \mathbb{R}^d$ is a standard Brownian motion and $\vec{v}(\vec{r}_t,t)$ is a drift that is determined by the potential $V(\vec{r})$, as well as the initial and final conditions. The problem of finding the drift can be given an optimal control formulation, with $\vec{v}(\vec{r}_t,t)$ achieving the minimum of the cost function (\cite{Holland:1977aa,Fleming:1977aa})
\begin{equation}\label{eq:hollandcost}
  C_T[\vec{v}] = \frac{1}{T}\E\left[\int_0^T\left[\frac{1}{2}(\vec{v}(\vec{r}_t,t))^2 + V(\vec{r}_t)\right]dt\right],
\end{equation}
where the expectation is over the process \eqref{eq:sde}. When this cost is minimized the path measure of this process coincides with the FK measure. For the infinite horizon $T\to\infty$ case the optimal drift has no explicit time dependence and we obtain the ground state energy as $E_0 = \lim_{T\to\infty} \min_{\vec{v}} C_T[\vec{v}]$.

The existence of a variational principle allows the tools of deep learning to be brought to bear. In this work the drift function will be parameterized by a neural network with parameters $\theta$: $\vec{v}(\vec{r})=\vec{v}_\theta(\vec{r})$. Finding the optimal drift can be regarded as a \emph{reinforcement learning} problem with the cost \eqref{eq:hollandcost} as our (negative) reward function. Although the expectation in \eqref{eq:hollandcost} is intractable, a Monte Carlo estimate may be made by generating a batch of solutions of the SDE using standard discretizations (\cite{Kloeden:2013aa}). Optimization of this estimate is then possible through automatic differentiation of the estimated cost with respect to $\theta$. Derivatives with respect to the expectation are straightforward as the SDE increments are written in terms of standard Brownian increments, analogous to the reparameterization trick used in stochastic backpropagation through deep latent Gaussian models (\cite{Rezende:2014aa}) or variational autoencoders (\cite{Kingma:2013aa}).


As well as providing a route to the evaluation of the ground state energy, \eqref{eq:sde} provides a sampler for the ground state probability distribution, as the stationary distribution of the SDE with the optimal drift coincides with $|\varphi_0|^2$.

The imaginary time Feynman path integral (or FK formula) is the basis of other numerical techniques in physics, notably the path integral Monte Carlo method (\cite{Ceperley:1995aa}). In this approach, the state of the Monte Carlo simulation corresponds to an entire Feynman trajectory, which is updated according to a Markov chain that ensures the paths sample the FK distribution in the stationary state. Our approach by contrast learns to sample the trajectories in the optimal way. We intend our method to be a drop-in replacement for path integral Monte Carlo as far as ground state properties are concerned.

\subsection{The Many-Body Problem}

All of the above considerations extend straightforwardly to systems of many identical particles, with one important caveat. The wavefunction of a system of identical particles must either be completely \emph{symmetric} or \emph{antisymmetric} under exchange of particles, corresponding to identical bosons or fermions (the latter described by the Pauli exclusion principle). The overall ground state of an identical particle Hamiltonian is symmetric under very general conditions and therefore corresponds to the bosonic case (\cite{Feynman:1998aa}). Apart from some two-electron problems where the spin state of the electrons is antisymmetric, allowing the spatial wavefunction to be symmetric (see \Cref{sec:exp}), the many-fermion problem is out of reach of our method for now.\footnote{The same restriction applies to the path integral Monte Carlo method, where it is known as the \emph{sign problem}.}

\subsection{Outline}

The outline of the remainder of this paper is as follows. In \Cref{sec:oc} we discuss the connection between quantum mechanics and optimal control in more detail, focusing on the variational principle that can be used to learn the optimal drift. In \Cref{sec:rl} we introduce our neural representation of the drift, with particular attention given to permutation equivariance which is a feature of systems of indistinguishable particles. We also describe the  algorithm used to learn the drift. \Cref{sec:exp} describes experiments on some simple physical systems. Finally, in \Cref{sec:con} we summarize our findings and provide an outlook to future work.

\section{Quantum Mechanics and Optimal Control}\label{sec:oc}


\subsection{Fokker--Planck to Schr\"odinger}\label{sec:fp}

The most direct way to establish a link between stochastic processes and quantum mechanics is via a standard mapping between the Fokker--Planck and Schr\"odinger equations. This is discussed for example in \cite{Risken:1996aa,Pavliotis:2014aa}, but we repeat the main points here. Starting from a solution $ p({\vec{r}},t)$ of the Fokker--Planck (FP) equation
\begin{equation} \label{eq:fp}
  \frac{\partial p({\vec{r}},t)}{\partial t} = \frac{1}{2} \nabla^2 p({\vec{r}},t) + \nabla \cdot \left ( p({\vec{r}},t) \nabla U({\vec{r}}) \right ),
\end{equation}
with a drift $\vec{v}(\br) = - \nabla U({\vec{r}})$ given in terms of some potential function $U({\vec{r}})$, the stationary state $\pi({\vec{r}})$ of this FP equation has the form of a Boltzmann distribution
\begin{equation}
  \pi({\vec{r}}) \propto \exp \left(-2 U({\vec{r}}) \right).
\end{equation}
If we define the function
\begin{equation}\label{eq:psi_def}
  \psi({\vec{r}},t) = \frac{p({\vec{r}},t)}{\sqrt{\pi({\vec{r}})}},
\end{equation}
then $\psi({\vec{r}},t)$ satisfies the (imaginary time) Schr\"odinger equation
\begin{equation}
  \frac{\partial \psi({\vec{r}},t)}{\partial t} = -H \psi({\vec{r}},t)
\end{equation}
with Hamiltonian $H=\frac{1}{2} \nabla^2+V_U(\vec{r})$,
%
%
where the potential $V_U(\vec{r})$ has the form
\begin{equation}\label{eq:FP-pot}
V_U(\vec{r})\equiv \frac{1}{2} \left[-\nabla^2 U + (\nabla U)^2\right]=\frac{1}{2}\left[\nabla\cdot \vec{v}+\vec{v}^2\right].
\end{equation}
The zero energy ground state wavefunction of this Hamiltonian is
\begin{equation}
  \varphi_0({\vec{r}}) = \sqrt{\pi({\vec{r}})}.
\end{equation}
The Fokker--Planck equation \eqref{eq:fp} describes the evolution of the probability density of a stochastic process described by the stochastic differential equation (SDE)
\begin{equation}\label{eq:SDE2}
  d{\vec{r}}_t = d\vec{B}_t +\vec{v}({{\vec{r}}_t})dt,
\end{equation}
where $\vec{B}_t$ is a standard Brownian motion. 
We emphasize that the quantum probability in the ground state $|\varphi_{0}({\vec{r}})|^2$ coincides with the classical stationary distribution $\pi_0$ of this process. 

\subsubsection{Example: Calogero--Sutherland and Dyson Brownian Motion}

An instructive example of a many body problem and the associated stochastic process is provided by the Calogero--Sutherland model describing particles in one dimension in a harmonic potential interacting by an inverse square potential (\cite{Sutherland:1972aa})
$$
H = \sum_i \frac{1}{2}\left[-\frac{\partial^2}{\partial x_i^2}+x_i^2\right] + \lambda(\lambda-1)\sum_{i<j} \frac{1}{(x_i-x_j)^2}.
$$
In this case, the ground state is known exactly and has the form of the Gaussian ground state wavefunction of the harmonic oscillator multiplied by a pairwise Jastrow factor
$$
\Phi_0(x_1,\ldots x_N) = \prod_{i<j}|x_i-x_j|^{\lambda}\exp\left(-\frac{1}{2}\sum_i x_i^2\right).
$$
The drift of the associated SDE is
\begin{equation}\label{eq:CS_drift}
v_i =\partial_i \log\Phi_0= - x_i + \lambda \sum_{j\neq i} \frac{1}{x_i-x_j}
\end{equation}
corresponding to Dyson's Brownian motion (\cite{Dyson:1962aa}, see \Cref{fig:cs-sim}).

\begin{figure}
\floatconts
  {fig:cs-sim}
  {\caption{Trajectories of 15 particles evolving under the Dyson process \eqref{eq:CS_drift}.}}
  {\includegraphics[width=0.5\textwidth]{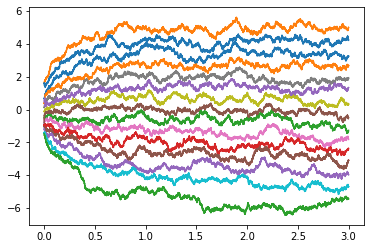}}
\end{figure}

\subsection{The Variational Principle}\label{sec:var}

In order to turn the above connection into a calculational tool, we exploit a connection between the path measure $\mathbb{P}_{\bv}$ of the SDE \eqref{eq:SDE2} and the corresponding quantity $\mathbb{P}_\text{FK}$ from the FK formula. For a FP measure that relates the ground state distribution $\pi_0(\br_{0,T})$ at $t=0$ and $t=T$, the log likelihood ratio (log Radon--Nikodym derivative) of the two measures obeys
\begin{align}\label{eq:KLlim}
  \log\left(\frac{d\mathbb{P}_{\bv}}{d\mathbb{P}_\text{FK}}\right) &=\ell_T - E_0 T+\log\left(\frac{\varphi_0(\vec{r}_0)}{\varphi_0(\vec{r}_T)}\right)\nonumber\\
   \ell_T&\equiv \int \bv(\br_t)
  \cdot d\vec{B}_t+\int dt\left(\frac{1}{2}|\bv(\br_t)|^2+V(\br_t)\right),
\end{align}
See Appendix \ref{sec:prob-proof} for a heuristic proof. The  Kullback–Leibler (KL) divergence of the two measures is then
\begin{equation}\label{eq:full-KL}
    D_{\text{KL}}\left(\mathbb{P}_{\bv} \middle\| \mathbb{P}_\text{FK}\right)=\E_{\mathbb{P}_{\bv}}\left[\ell_T-E_0 T+\log\left(\frac{\varphi_0(\vec{r}_0)}{\varphi_0(\vec{r}_T)}\right)\right].
\end{equation}
Note that \eqref{eq:full-KL} is true for any $\vec{r}_0$. If we additionally average over $\vec{r}_0$ drawn from the stationary distribution of the SDE, then the distributions of $\vec{r}_T$ and $\vec{r}_0$ coincide, and the final term vanishes. Because the KL divergence is positive $D_{\text{KL}}\left(\mathbb{P}_{\bv} \middle\| \mathbb{P}_\text{FK}\right)\geq 0$ we then have $\E_{\mathbb{P}_{\bv}}\left[\ell_T\right]\geq E_0 T$, with equality when the two measures match. This is the variational principle that forms the basis of our approach. For $T\to\infty$ we can ignore the final term so that
\begin{equation}
    \lim_{T\to\infty} \frac{\E_{\mathbb{P}_{\bv}}\left[\ell_T\right]}{T}\geq E_0,
\end{equation}
irrespective of the initial state distribution (as $T\to\infty$ the final state distribution will be the stationary state of the SDE, assuming ergodicity). Using the fact that Brownian increments have zero expectation recovers Holland's cost \eqref{eq:hollandcost}.

\section{Reinforcement Learning}\label{sec:rl}

We have identified a variational principle for the drift with a time integrated cost function. Learning the optimal drift may be regarded as an entropy regularized reinforcement learning (RL) problem. The usual Markov decision process formulation of RL involves Markov dynamics in a state space together with a set of actions that can influence the dynamics within this space. We seek a \emph{policy} -- a distribution of actions to be taken at each step given the current state -- optimal with respect to a time integrated reward function. In the present case, optimizing over all possible drift functions corresponds to complete controllability of the state, instead of a restriction to a (possibly finite) set of actions. As we have seen in \Cref{sec:var}, however, the optimization problem is regularized by a KL divergence that originates from the kinetic energy, and penalizes deviations of the stochastic process from the `passive' Brownian dynamics. In RL the resulting cost gives rise to \emph{linearly solvable Markov decision problems} (\cite{Todorov:2007aa,Todorov:2008aa}). \footnote{In RL this entropy regularization has a practical utility in encouraging diverse policies (\cite{Levine:2018aa}).}

In the remainder of this section we introduce a concrete algorithm to learn the optimal drift for a given quantum many-body problem.

\subsection{Neural representation of the drift}\label{sec:drift}

In practice, one cannot optimize over arbitrary drifts. We introduce a neural parameterization $\vec{v}_\theta(\vec{r})$ of the drift function, so the quality of our approximation is limited by the expressiveness of the network. Thus we are performing a variational approximation to the optimal control problem, with the cost bounded from below by the ground state energy.   

For $N>1$ particles, the drift of each is a function of all the particle coordinates: $\vec{v}_{i,\theta}(\vec{r}_1,\ldots,\vec{r}_N)$ $i=1,\ldots, N$. For identical particles, this drift must be \emph{permutation equivariant}, meaning that for any permutation $P$ of the particles
\begin{equation}
    \vec{v}_{i,\theta}(\vec{r}_1,\ldots,\vec{r}_N) = \vec{v}_{P(i),\theta}(\vec{r}_{P(1)},\ldots,\vec{r}_{P(N)}).
\end{equation}
We now introduce two ways in which equivariance may be achieved.

\subsubsection{DeepSets}\label{sec:deepsets}

\cite{Guttenberg:2016aa,Zaheer:2017aa} identified deep neural architectures that respect permutation equivariance, building on the much earlier work of \cite{Shawe-Taylor:1989aa}. We use a variant of the \textsc{DeepSets} architecture of \cite{Zaheer:2017aa}, in which an equivariant hidden state is based on the sum-pooling operation
$$
\mathbf{h}_i = \phi\left(\boldsymbol{\Gamma}\br_i + \boldsymbol{\Lambda}\sum_j\br_j\right)
$$
(biases omitted for clarity) for weight matrices $\boldsymbol{\Gamma}, \boldsymbol{\Lambda}\in \mathbb{R}^{H\times d}$, where $d$ is the spatial dimension and $H$ the number of hidden units. $\phi$ is a nonlinear activation function. These layers can then be stacked to produce a more expressive representation.

\subsubsection{PairDrift}\label{sec:pairdrift}

Alternatively, we may use a representation of single particle and pair features (c.f. \cite{Pfau:2019aa})
\begin{align}
\mathbf{h}_i &= \boldsymbol{\sigma}_1(\mathbf{r}_i) + \sum_j \boldsymbol{\pi}_1(\mathbf{r}_i-\mathbf{r}_j)\\
\mathbf{h}_{ij} &= \boldsymbol{\Pi}_1(\mathbf{r}_i-\mathbf{r}_j).
\end{align}
where $\boldsymbol{\sigma}, \boldsymbol{\pi}:\mathbb{R}^d\to \mathbb{R}^H$ and $\boldsymbol{\Pi}:\mathbb{R}^d\to \mathbb{R}^{H\times H}$ are neural networks. The single particle hidden state $\mathbf{h}_i$ contains information about single particle and pair locations. This process may be continued
\begin{align}
\tilde{\mathbf{h}}_i &= \boldsymbol{\sigma}_2(\mathbf{h}_i) + \sum_j \boldsymbol{\pi}_2(\mathbf{h}_{ij})\\
\tilde{\mathbf{h}}_{ij} &= \boldsymbol{\Pi}_2(\mathbf{h}_{ij}).
\end{align}
Finally, the permutation equivariant drift function is written in terms of the last hidden state as
\begin{equation}
\bv_i = \boldsymbol{\sigma}_3(\tilde{\mathbf{h}}_i) + \sum_j \boldsymbol{\pi}_3(\tilde{\mathbf{h}}_{ij}).
\end{equation}
Note that if we had represented the drift simply as
$$
\bv_i = \boldsymbol{\sigma}(\mathbf{r}_i) + \sum_j \boldsymbol{\pi}(\mathbf{r}_i-\mathbf{r}_j),
$$
this would correspond to a pairwise Jastrow factor in the many-body wavefunction via the formula $\bv_i=\nabla_i \log\Psi$. Repeatedly merging the pair features into the single particle features enriches the representation. In our experiments, we find it sufficient to include a single hidden layer. 

\subsection{Integration of the SDE} \label{sec:int}

From \eqref{eq:KLlim}, we see that to evaluate the cost function we need to solve the system of SDEs
\begin{align}\label{eq:sde_system}
  d\vec{r}_t &= d\vec{B}_t + \vec{v}(\vec{r}_t)dt,\nonumber\\
  d\ell_t &= \vec{v}(\vec{r}_t)\cdot d\vec{B}_t + \left(\frac{1}{2}|\vec{v}(\vec{r}_t)|^2 + V(\vec{r}_t)\right)dt.
\end{align}
As $T\to\infty$, $\ell_T$ satisfies $\E_{\mathbb{P}_{\bv}}\left[\ell_T\right]\geq E_0 T$. Numerous techniques exist for the numerical solution of such problems, beginning with the Euler--Maruyama scheme
\begin{align}\label{eq:em}
    \vec{r}_{t+\Delta t} &= \vec{r}_{t} + \Delta\vec{B}_t + \vec{v}(\vec{r}_t)\Delta t,\nonumber\\
    \ell_{t+\Delta t} &= \ell_{t} + \vec{v}(\vec{r}_t)\cdot \Delta\vec{B}_t +\left(\frac{1}{2}|\vec{v}(\vec{r}_t)|^2 + V(\vec{r}_t)\right)\Delta t.
\end{align}
where $\Delta\vec{B}_t\sim \mathcal{N}(0,t)$ are iid Gaussian increments. 

Higher order methods offering improved accuracy with longer time steps $\Delta t$ are described in \cite{Kloeden:2013aa}. For most of our experiments we used the SOSRA method from \cite{Rackauckas:2018aa}. Implementing these methods in a framework that supports automatic differentiation allows the cost to be optimized by gradient descent with respect to the parameters $\theta$ describing the drift.

As emphasized in recent work on neural ordinary differential equations \cite{Chen:2018aa}, one can regard integration schemes like \eqref{eq:em} as a (recurrent) residual network. This affords a clean separation of the model (the drift $\vec{v}_\theta(\vec{r})$) from the implementation and allows 
us to adapt the integration scheme on the fly for different epochs of training, or at evaluation time.

We evolve a batch of trajectories for $T/\Delta t$ steps of size $\Delta t$, starting from an initial state equal to the final state of the previous batch. This is analogous to training a recurrent network by backpropagation through time on infinite sequences by unrolling for a finite number of steps, stopping the gradients from propagating into the infinite past. Starting each step from the final state of the previous step means that the batch tracks closely the stationary distribution of the stochastic process, so no burn-in is required before the integration of the cost function can begin. 

This procedure leads to the following issue, arising from the presence of the boundary term in \eqref{eq:KLlim}. Its expectation is zero for a stationary process where the marginal distributions of the initial and terminal points coincide. The gradients of this expectation with respect to the drift parameters are however nonzero, as changing these parameters changes the endpoint $\vec{r}_T$. Evaluating this gradient to get the gradient of the KL in \eqref{eq:full-KL} is intractable because the ground state wavefunction is not known, but it may be estimated from the current drift as
\begin{equation}\label{drift-correct}
\nabla_\theta \log\left(\frac{\varphi_0(\vec{r}_0)}{\varphi_0(\vec{r}_T)}\right) = -\nabla_\theta \vec{r}_T \cdot \vec{v}(\vec{r}_T)\approx -\nabla_\theta \vec{r}_T \cdot \vec{v}_\theta(\vec{r}_T),
\end{equation}
an approximation that improves as the true drift $\vec{v}(\vec{r})=\nabla\log\varphi_0(\vec{r})$ is approached. Although this correction may be neglected in the $T\to\infty$ limit, we find it significantly improves convergence in our experiments.

\subsection{Stochastic Backpropagation}

Recall that the cost function \eqref{eq:full-KL} is the expectation of a time integral over the stochastic process. Although the expectation is intractable, a batch of $B$ simulated trajectories $\vec{r}^{(b)}_{t}$ can be used to make a Monte Carlo estimate as follows 
\begin{equation}\label{eq:cost-est}
    \ell_T[\vec{v}_\theta] \approx \frac{1}{B T} \sum_{b,t}\left[\vec{v}_\theta\left(\vec{r}^{(b)}_t\right)\cdot \Delta\vec{B}^{(b)}_t + \left(\frac{1}{2}\left(\vec{v}_\theta\left(\vec{r}^{(b)}_t\right)\right)^2 + V\left(\vec{r}^{(b)}_t\right) \right)\Delta t \right ].
\end{equation}
Because $\vec{r}^{b}_{t}$ have been generated by an SDE discretization such as \eqref{eq:em}, backpropagating through this estimated expectation is analogous to the reparameterization trick in deep latent Gaussian models (\cite{Rezende:2014aa}) or variational autoencoders (\cite{Kingma:2013aa}). This similarity has been emphasized in recent work on neural stochastic differential equations (\cite{Tzen:2019aa}). The memory cost of backpropagation is proportional to $BT/\Delta t$, and is the main bottleneck in our approach. 

Though the exact $T\to\infty$ value of the cost is an upper bound on the ground state energy, we have to bear in mind that the approximations of finite time, finite batch size, and any discretization error in the integration of the SDE mean that in practice the upper bound may be violated during training, particularly as optimization converges.



\subsection{Related Work}

We have already mentioned wavefunction based neural approaches to quantum mechanics in \Cref{sec:dlq}. Here we summarize work that is technically related to ours.

Learning SDEs using stochastic optimization was introduced in \cite{Ryder:2018aa} to infer SDE parameters from trajectories and in \cite{Tzen:2019aa} as a generative model to map a simple initial distribution of $\vec{r}_{-T/2}$ to a given empirical distribution of $\vec{r}_{T/2}$.

The path integral point of view has been applied to RL in \cite{Kappen:2007aa,Theodorou:2010aa}. Also from the RL side, \cite{Todorov:2007aa} introduced \emph{linearly solvable Markov decision problems}. Their reward function -- as in the quantum case -- is a sum of a cost on the state space and the KL divergence between the passive and controlled dynamics. Todorov showed that for these problems the nonlinear Bellman equation could be transformed to a linear equation by an exponential transformation. As we have seen in \Cref{sec:oc}, for the quantum cost this linear equation is just the Schr\"odinger equation. 

Finally, \cite{Han:2018aa} solved semilinear parabolic PDEs in high dimension using a representation in terms of a backward SDE with a neural representation, but with a cost function based on matching to a given terminal condition.

\section{Experiments} \label{sec:exp}

Our experiments use \textsc{PairDrift} architectures described in \Cref{sec:drift}, which we found to perform best, with a single hidden layer and HardTanh activation. To ensure the trajectories remain close to the origin from the outset, these networks are initialized to give zero drift, but we add a skip connection to give a restoring drift of simple form. For example, a linear skip connection $\mathbf{W}_\text{skip}\br_i$ to the drift of the $i^\text{th}$ particle ensures that the SDE is initially a 3D Ornstein--Uhlenbeck process.

Optimization of the estimated cost \eqref{eq:cost-est} uses Adam (\cite{Kingma:2015ab}). The gradient is corrected to account for the boundary terms as described in \Cref{sec:int}
\begin{equation}
    \nabla_\theta \ell_T \to \nabla_\theta \ell_T-\nabla_\theta \vec{r}_T \cdot \vec{v}_\theta(\vec{r}_T)
\end{equation}
All models were implemented in PyTorch. For hyperparameters see Appendix \ref{sec:hyperparameters}. The code is available at \url{https://github.com/AustenLamacraft/QuaRL}

\subsection{Hydrogen Atom}


    

Hydrogen is an electrically neutral atom containing a nucleus consisting of a single proton, and a single electron which is bound to the nucleus via the Coulomb force. Consider a single non-relativistic electron subject to an isotropic Coulomb interaction force provided by the nucleus. In atomic units\footnote{\emph{Atomic units} denotes energies measured in Hartrees (27.211 eV) and distances measured in Bohr radii (0.529 $\AA$).} the explicit form of the Hamiltonian reads
\begin{equation}\label{eq:hydrogen_Hamiltonian}
H = -\frac{1}{2}\nabla^{2} - \frac{1}{|\br|}
\end{equation}
where $\br$ is the position of the electron with respect to the center of the nucleus. The Schr\"odinger equation for hydrogen can be solved explicitly via separation of variables, yielding the ground state wave-function
\begin{equation}\label{eq:hydrogen_wave} 
\varphi_{0}(\br) = \frac{1}{\sqrt{\pi}}e^{-|\br|}.
\end{equation}
As the ground state of hydrogen has an exact analytic form, it is also possible to derive an exact expression for the electron's drift. The ground state drift velocity of the electron is given by
\begin{equation}\label{eq:hydrogen_drift}
\bv = \nabla\log \varphi_0(\br) = -\hat{\br}.
\end{equation} 
The exact ground state energy of Hydrogen is $E_{0}= -0.5$. We used a skip connection with $\mathbf{W}_\text{skip}=-\mathsf{1}_3$, and found an energy of $-0.497$, which is within $0.6 \%$  of the exact value. The learned drift is illustrated in \Cref{fig:h-and-he}. From \eqref{eq:hydrogen_drift} and Figure \ref{fig:h-and-he}, we see that the Hydrogen ground state drift has a cusp at the position of the proton.

\begin{figure}[htbp]
\floatconts
  {fig:h-and-he}
  {\caption{(Left) The learned drift field for the Hydrogen atom in the $x-y$ plane, colorscaled by its magnitude. (Right) The electron density distribution in the stretched $\text{H}_2$ molecule at proton separation $R=2.8$.}}
  {\includegraphics[height=0.43\textwidth]{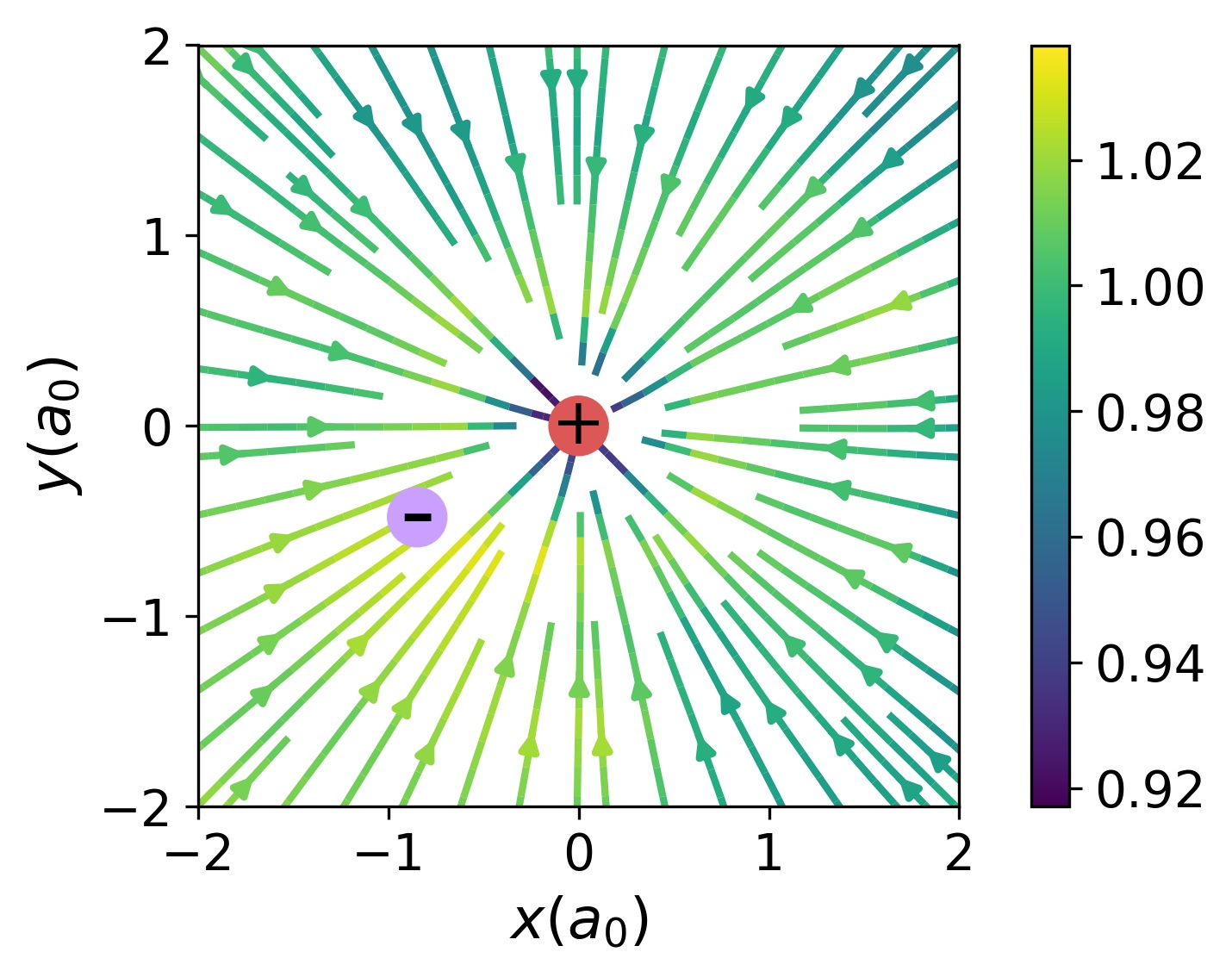}\hfill\includegraphics[height=0.43\textwidth]{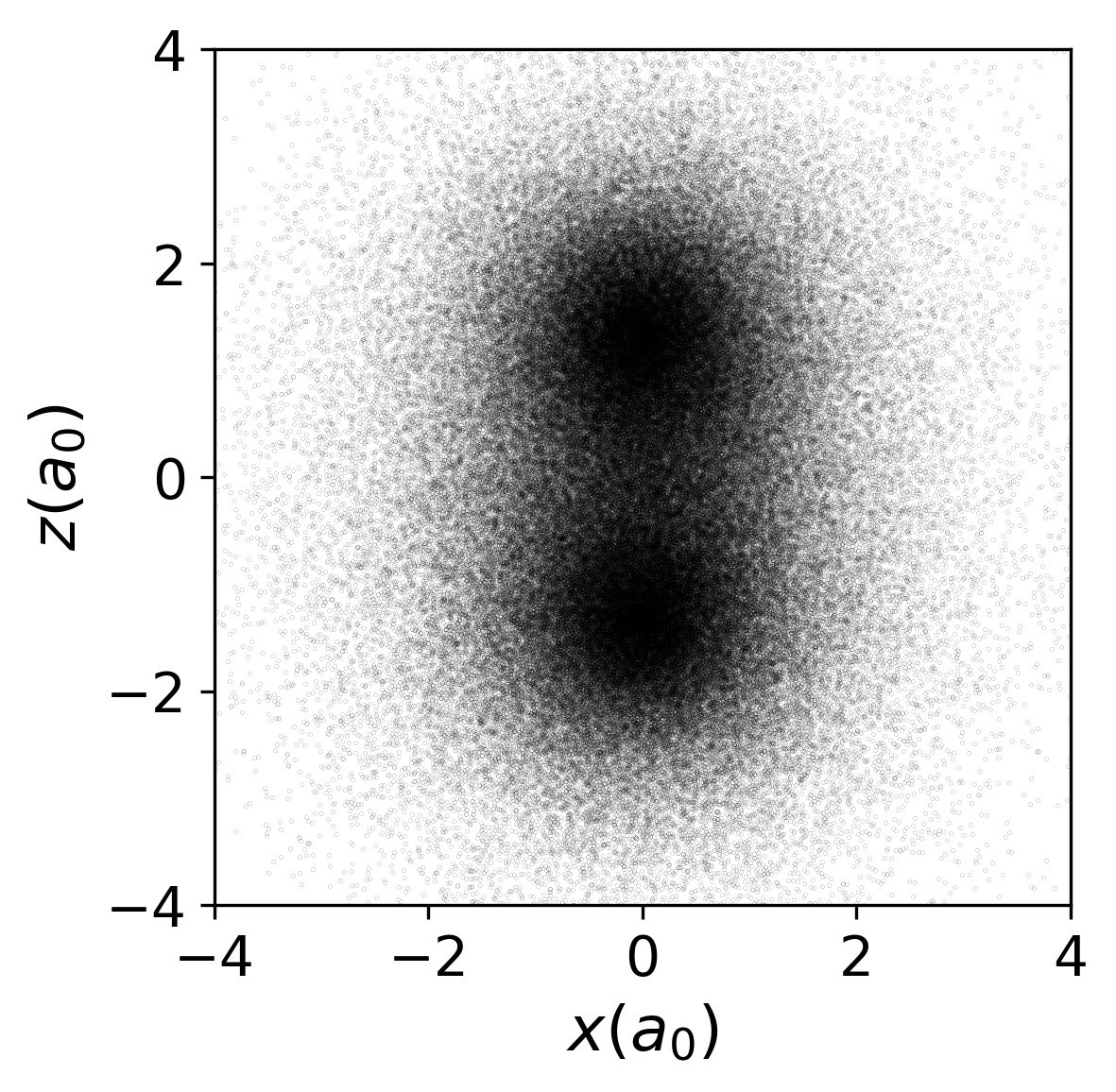}}
\end{figure}

\subsection{Helium atom and Hydrogen molecule}
The Helium atom consists of two electrons bound to a nucleus of charge $+2e$. The non-relativistic two-body Hamiltonian reads
\begin{equation}\label{eq:helium_atom_Hamiltonian} H = - \frac{1}{2}    \left(\nabla_{\br_1}^{2} + \nabla_{\br_2}^{2} \right ) -  \frac{2}{|\br_1|} - \frac{2}{|\br_2|} + \frac{1}{|\br_{1} - \br_{2}|} 
\end{equation} 
where $\br_{1}$ and $\br_{2}$ denote the positions of the electrons with respect to the  nucleus. An analytic solution to the Schr\"odinger equation does not exist for Helium. Accurate numerical calculations may be found in the recent paper \cite{Baskerville:2019aa}, together with a comparison with the Hartree--Fock method. In the Hartree--Fock (HF) approximation the wavefunction is taken as the product of two single particle wave-functions \footnote{The antisymmetry of the wavefunction demanded by the Pauli principle is achieved for the ground state by putting the spin states of the two electrons in a spin singlet. Spin triplet states -- requiring an antisymmetric spatial wavefunction -- are not accessible by our method.}
\begin{equation}\label{eq:helium_wave} 
\Psi_{\text{HF}}\left(\br_1 , \br_2\right) = \psi_{\text{HF}}\left(r_1\right)\psi_{\text{HF}}\left(r_2\right),
\end{equation}
where the single particle wavefunction $\psi_\text{HF}(r)$ is optimized to minimize the energy. \cite{Baskerville:2019aa} find a correlation energy -- the difference between the best HF value and the numerically exact values -- of $0.042$, or $1.4 \%$ of the exact value.

We used the \textsc{PairDrift} network to model the Helium atom. The ground state energy was found to be within $0.2\%$ of the exact value. 

For both the Helium atom and the Hydrogen molecule, we used a skip connection taking advantage of Kato's cusp conditions (\cite{Kato:1957aa}). These are conditions on the cusps that appear in the wavefunction whenever two particles with a Coulomb interaction are at the same position. Kato's cusp conditions for wave-functions translate directly into cusp conditions for drifts: they should be proportional to $\hat{\br}$ in the limit of zero separation. For electron-electron interactions, the proportionality factor is $\frac{1}{2}$ and the drift is repulsive, while for electron-nucleus interactions, the proportionality factor is the charge of the nucleus and the drift is attractive. For our experiments on the Helium atom and Hydrogen molecule, we therefore use a skip connection that is a sum of terms $\hat{\br}$, i.e. one term for each Coulomb interaction.

Although we could have incorporated other conditions arising from rotational symmetries, we chose not to do so because these symmetries do not generalize to other continuum systems.

The Hydrogen molecule consists of two electrons orbiting a pair of bound protons at separation $R$, which repel each other. With the protons at positions $\pm R\hat{\bz}/2$, the Hamiltonian becomes
\begin{equation}\label{eq:hydrogenMol_Hamiltonian}
H = - \frac{1}{2}   \left(\nabla_{\br_1}^{2} + \nabla_{\br_2}^{2} \right)  + \frac{1}{R} + \frac{1}{|\br_{1} - \br_{2}|} -  \frac{1}{|\br_{1} - R\hat{\bz}/2|} -  \frac{1}{|\br_{1} + R\hat{\bz}/2|}
-  \frac{1}{|\br_{2} - R\hat{\bz}/2|} -  \frac{1}{|\br_{2} + R\hat{\bz}/2|}.
\end{equation}
The correlation energy of the Hydrogen molecule is -0.044 Hartrees, corresponding to 3.8\% of the numerically exact value. 

To model the Hydrogen molecule, we also used the  \textsc{PairDrift} network with the skip connection describing the electron-electron and electron-nucleus cusps. We find a ground state energy within $0.3 \%$ of the exact value at the equilibrium separation of $R=1.401$. For a stretched Hydrogen molecule at $R=2.800$ (i.e. twice the equilibrium separation), our result is similarly $0.3 \%$ from the exact value. 

All our results for the atomic and molecular systems are given in \Cref{tab:results}.


\subsection{Bosons in a Harmonic Trap}

As an example of a many-particle system we study a system of $N$ identical bosons trapped in a two-dimensional isotropic harmonic potential, with the atom-atom interactions approximated by a finite-sized Gaussian contact potential (\cite{Mujal:2017aa}).  The relevant Hamiltonian reads
\begin{equation}\label{eq:harmonic_Hamiltonian} 
H =  \sum_{i = 1}^{N} \left[ - \frac{1}{2}  \nabla_{i}^{2} + \frac{1}{2}\vec{\br_{i}}^{2} \right] +  \frac{g}{\pi s^{2}}\sum_{i < j }^{N}\exp\left[-\left(\br_{i} - \br_{j}\right)^2/s^2\right] 
\end{equation} 
where $g$ and $s$ are respectively the strength and range of the pairwise interaction. \break

We simulated the ground state for $N = 2,3,$ and $4$ bosons for $s=0.5$ and $g=0-15$ using the \textsc{PairDrift} neural architecture. The energies derived via the neural network are compared with those of \cite{Mujal:2017aa} (obtained by exact diagonalization) in \Cref{fig:boson-drift-and-results}. An energy correspondence within 1\% of the exact diagonalization results was observed for all parameter combinations.
\break

\begin{figure}[htbp]
\floatconts
  {fig:boson-drift-and-results}

 \includegraphics[height=0.32\textwidth]{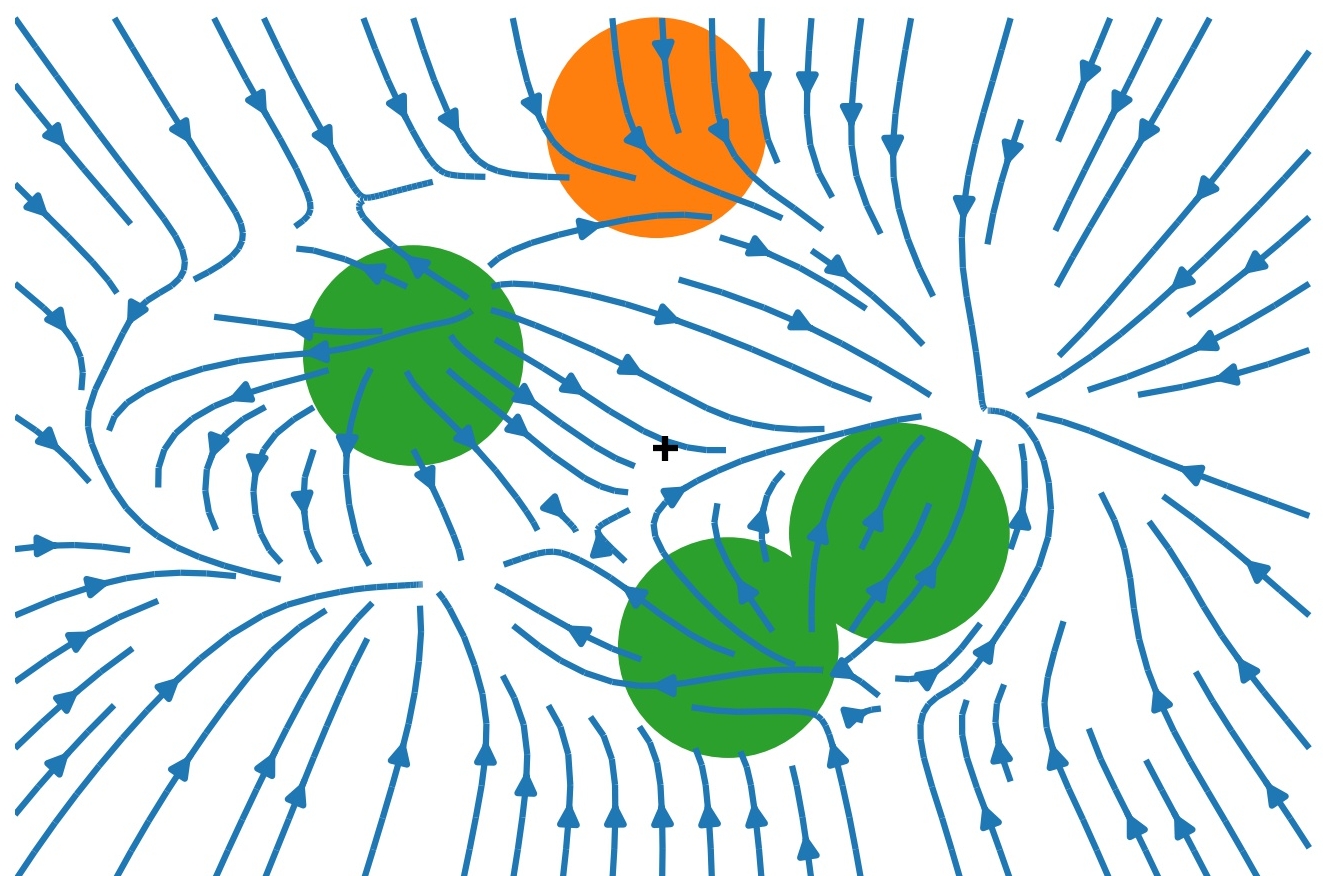}\hfill\includegraphics[height=0.32\textwidth]{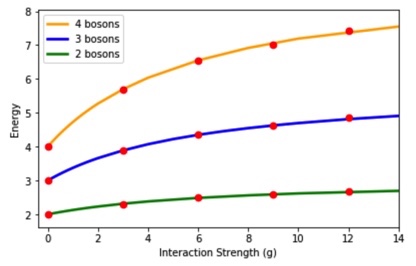}
 
 {\caption{(Left) The learned drift field for interaction strength $g=15$ of one boson (orange) in the presence of the others (green), together with the restoring drift arising from the harmonic potential pushing particles to the center (cross). Note the resultant drift field tends to keep the bosons apart. (Right) Comparison of ground state energies of $N=2$, $N=3$, and $N=4$ bosons (blue, orange and green, respectively) with results of \cite{Mujal:2017aa}.}}
 
\end{figure}

\begin{table}[]  
    \centering
    \begin{tabular}{ p{3cm} | p{2.3cm} p{2.3cm} p{2.3cm} p{2.3cm}} 
              & H atom    & He atom   & $\text{H}_2$ molecule & $\text{H}_2$ molecule ($R=2.8$)\\
 \hline
  Numerically exact  & -0.5      & -2.903    &  -1.173   & -1.071\\
 Hartree-Fock   & N/A        & -2.862 (1.4\%)   &  -1.129 (3.8\%)   & \\
 \textbf{Ours}  &-0.497 (0.6\%)  & -2.898 (0.2\%) &  -1.169 (0.3\%)    & -1.068 (0.3\%)
\end{tabular}
    \caption{Comparison of our approach with Hartree-Fock and numerically exact values for the ground state energies. The percentages within parentheses are the relative errors of an energy estimate compared to the numerically exact value.}
    \label{tab:results}
\end{table}

\hfill \break

\section{Conclusion}\label{sec:con}

In this work we have presented a novel neural approach to many-body quantum mechanics that leverages the optimal control formulation of \cite{Holland:1977aa} to approximate the Feynman--Kac path measure. This contrasts with earlier deep learning approaches to the many-body problem that start from the Schr\"odinger picture. We have demonstrated the utility of our approach for some simple many-body problems.

Future work will extend this approach in the following directions:

\begin{enumerate}
    \item Many models studied in condensed matter physics are defined on a \emph{lattice} rather than in continuous space. In this case identical particles are handled by working with \emph{occupation numbers} -- the count of particles on each lattice site -- rather than by introducing a fictitious index for each particle and then demanding invariance under labelling. The control theoretic perspective is well suited to such discrete state spaces as described in \cite{Todorov:2007aa}.
    \item Rotating systems or systems with finite angular momentum may be handled with a modified cost function that
    leads to a rotational component to the drift. This is how time reversal symmetry is broken in our formalism, corresponding to complex wavefunctions in the Schr\"odinger picture.
    
    \item The memory cost of backpropagating through SDE trajectories is a major limitation of our approach that could be addressed using checkpointing  \cite{Martens:2012aa,Chen:2016aa,Gruslys:2016aa}. \cite{Chen:2018aa} suggested that neural ODE models be trained at constant memory cost by solving the linear adjoint equation backwards in time. The difficulties associated with implementing this method in the SDE case are discussed in \cite{Tzen:2019aa}, though very recent work has addressed this issue \cite{li2020scalable}.
    
\end{enumerate}

\acks{We thank Alex Gaunt for useful discussions. We acknowledge support from a National Science Foundation Graduate Research Fellowship under Grant No. DGE-174530 (ARB) and EPSRC Grant No. EP/P034616/1 (AL).}

\bibliography{msml}
\appendix

\section{Probabilistic interpretation of the cost function} \label{sec:prob-proof}

We show that the KL divergence between the probability of paths under the stochastic process \eqref{eq:SDE2} and under the Feynman--Kac measure is, in the limit of long paths, given by \eqref{eq:KLlim}.

A very general Gibbs variational principle for functionals of Brownian motion, including functionals of FK type, appears in \cite{Boue:1998aa}. Here we provide only a heuristic argument adapted to the $T\to\infty$ limit of interest for the ground state properties.

We first consider the Radon--Nikodym (RN) derivative between $\mathbb{P}_{\bv}$ and the distribution of Brownian paths $\mathbb{P}_0$. Girsanov's theorem tells us that this is
\begin{equation}
  \frac{d\mathbb{P}_{\bv}}{d\mathbb{P}_0}=\exp\left(\int \bv(\br_t)\cdot d\br_t - \frac{1}{2}\int |\bv(\br_t)|^2 dt\right).
\end{equation}
Next, the Feynman--Kac formula can be expressed as the RN derivative (\cite{dai1990markov})
\begin{equation}\label{eq:RN2}
  \frac{d\mathbb{P}_\text{FK}}{d\mathbb{P}_0} = \frac{\tilde\psi(\vec{r}_T,T)}{\tilde\psi(\vec{r}_0,0)}\exp\left(-\int_{0}^{T} V(\vec{r}_t)dt\right),
\end{equation}
where $\tilde\psi(\vec{r}_t,t)$ is the solution of the `backwards' imaginary time Schr\"odinger equation
\begin{equation}
      \frac{\partial\tilde\psi(\vec{r},t)}{\partial t} = \left[H\tilde\psi\right](\vec{r},t).
\end{equation}
The RN derivative \eqref{eq:RN2} defines a path measure that relates initial and final distributions 
\begin{align}
    \pi(\br,0)&=\psi(\br,0)\tilde\psi(\br,0)\nonumber\\
    \pi(\br,T)&=\psi(\br,T)\tilde\psi(\br,T).
\end{align}
If we take both distributions to be the ground state distribution $\pi_0(\br)$ then
$$
\frac{\tilde\psi(\vec{r}_T,T)}{\tilde\psi(\vec{r}_0,0)}= e^{E_0T}\frac{\varphi_0(\vec{r}_T)}{\varphi_0(\vec{r}_0)}.
$$
Using these results to evaluate the overall RN derivative gives
\begin{align}\label{eq:RN-eval}
  \log\left(\frac{d\mathbb{P}_{\bv}}{d\mathbb{P}_\text{FK}}\right)&=\log\left(\frac{d\mathbb{P}_{\bv}}{d\mathbb{P}_0}\frac{d\mathbb{P}_0}{d\mathbb{P}_\text{FK}}\right)\nonumber\\
  &=\int \bv(\br_t)
  \cdot d\br_t+\int dt\left(-\frac{1}{2}|\bv(\br_t)|^2+V(\br_t)\right) - E_0 T + \log\left(\frac{\varphi_0(\vec{r}_0)}{\varphi_0(\vec{r}_T)}\right)\nonumber\\
  &=\int \bv(\br_t)
  \cdot d\vec{B}_t+\int dt\left(\frac{1}{2}|\bv(\br_t)|^2+V(\br_t)\right) - E_0 T + \log\left(\frac{\varphi_0(\vec{r}_0)}{\varphi_0(\vec{r}_T)}\right)\nonumber\\
  &=\ell_T - E_0 T + \log\left(\frac{\varphi_0(\vec{r}_0)}{\varphi_0(\vec{r}_T)}\right)
\end{align}
This yields the result \eqref{eq:KLlim}.

It is instructive to show how the RN derivative vanishes when the drift is optimal. In this case $\vec{v}_U(\br) = - \nabla U({\vec{r}})$ where $\varphi_0(\br)=e^{-U(\br)}$. It\^o calculus then gives
\begin{align}
dU(\vec{r}_t) = -v_U(\vec{r}_t)\cdot d\vec{r}_t -\frac{1}{2}\nabla\cdot \vec{v}_U(\vec{r}_t)dt,
\end{align}
so that
$$
U(\vec{r}_T)-U(\vec{r}_0)= - \int \vec{v}_U(\vec{r}_t)\cdot d\vec{r}_t-\frac{1}{2}\nabla\cdot \vec{v}_U(\vec{r}_t) dt.
$$
Thus the RHS of \eqref{eq:RN-eval} is
\begin{equation}\label{eq:ell-alt}
  \log\left(\frac{d\mathbb{P}_{\bv}}{d\mathbb{P}_\text{FK}}\right)=\int_0^T\left[V(\vec{r}_t)-V_U(\vec{r}_t) \right]dt - E_0 T,
\end{equation}
where $V_U=\frac{1}{2} \left(\nabla\cdot \vec{v}_U+\vec{v}_U^2\right)$ as before. This vanishes exactly when $V=V_U+E_0$ \emph{for any} $T$. This is the analogue of the `zero variance property' in variational quantum Monte Carlo (\cite{Foulkes:2001aa}) that greatly enhances the efficiency of that algorithm.

\section{Hyperparameters}
\label{sec:hyperparameters}

\begin{table}[]
    \centering
    \begin{tabular}{ c | c c c c c} 
                    & H atom    & He atom   & $\text{H}_2$ molecule & $\text{H}_2$ molecule ($R=2.8$) & Bosons\\
 \hline
 Batch size     & $2^{10}$        & $2^{10}$    &  $2^{10}$   &  $2^{10}$ & $2^{9}$\\
 Initial learning rate  & $10^{-2}$ & $10^{-3}$ & $5 \times 10^{-4}$ & $10^{-2}$ & $10^{-2}$ \\
 Number of time steps    & $2^{10}$ & $2^{10}$ & $2^{10}$ & $2^{10}$ & $2^{6}$\\
 Time step   & 0.01 & 0.01 & 0.01 & 0.01 & 0.01\\
 Width of hidden layer  & 256 & 64 & 64 & 64 & 64
\end{tabular}

    \caption{Hyperparameters used in training.     \label{tab:hyperparameters} }
\end{table}

We used an exponential decay of the learning rate: the learning rate was multiplied by a factor $0.95$ every $10$ training steps. For the bosons in a harmonic trap, we used the same hyperparameters for all simulations. The initial learning rates, as well as other training hyperparameters can be found in Table \ref{tab:hyperparameters}.





\end{document}